\documentclass[aps, pra, reprint, superscriptaddress, showkeys, floatfix]{revtex4-2}
\usepackage{bm}
\usepackage{graphicx}
\usepackage{amsmath,bm}
\usepackage{mathtools}
\usepackage[colorlinks=true,citecolor=blue,urlcolor=blue,linkcolor=blue]{hyperref}
\usepackage{overpic}
\begin{document}

\title{Quantum dynamics of a charged particle on rotating curved surfaces and space curves in electromagnetic fields}

\author{Long Du}
\email{ducoollong@qq.com}
\affiliation{Guangxi University of Science and Technology, Liuzhou 545006, China}

\author{Yong-Long Wang}
\email{wangyonglong@lyu.edu.cn}
\thanks{Corresponding author.}
\affiliation{School of Physics and Electronic Engineering, Linyi University, Linyi 276000, China}
\affiliation{Department of Physics, Nanjing University, Nanjing 210093, China}
\affiliation{National Laboratory of Solid State Microstructures, Nanjing University, Nanjing 210093, China}

\begin{abstract}
We develop a gauge-consistent framework for a charged particle confined to a rotating curved manifold in an electromagnetic field, starting from the covariant Klein--Gordon equation.
The non-relativistic reduction casts the rotational effects as effective gauge potentials while preserving the $U(1)$ symmetry, so that rotation and the external field enter alike, the geometry entering separately as the da Costa potential.
The obstruction to dimensional reduction is gauge--geometry entanglement: transverse magnetic fields on a two-dimensional (2D) surface and structural torsion on a one-dimensional (1D) curve couple the tangential dynamics to the transverse degrees of freedom.
We remove this entanglement by three mechanisms---a generalized transverse Poincar\'{e} gauge, the rotation-minimizing Bishop frame, and the parity of the transverse ground state, the last of which annihilates the residual first-order magnetic coupling for a single inversion-symmetric well.
On a torus, an axial field generates a joint Sagnac--Aharonov--Bohm residual of tunable sign without spoiling the poloidal--azimuthal separability, whereas an in-plane field couples only through the out-of-plane extent of the structure: it breaks the axial symmetry there but leaves a planar curve unaffected to first order.
The same torus shows the axial field tuning the curvature-induced localization, and a corrugated quantum ring converts periodic geometric and flux modulations into an artificial superlattice whose in-plane-field immunity follows from planarity and parity, not from the smallness of the field.
Non-inertial quantum dynamics on surfaces and curves of arbitrary shape is thus reduced, within the thin-layer regime, to two ingredients: an effective gauge potential and a geometric potential.
\end{abstract}


\maketitle

\section{Introduction}

The equivalence between non-inertial fictitious forces and classical electromagnetic fields has long served as a unifying framework for quantum mechanics in rotating reference frames \cite{anandan1977gravitational, mashhoon1988neutron}.
Quantum mechanics in non-inertial frames demands a gauge-consistent formulation, as macroscopic rotation modulates quantum interference to induce measurable phase shifts.
A direct coordinate transformation of the non-relativistic Schr\"{o}dinger equation in general destroys the manifest $U(1)$ gauge covariance, which can then be restored only by supplying the transformation law of the four-potential separately.
A consistent treatment therefore requires a covariant derivation from the Dirac or Klein--Gordon (KG) equations \cite{hehl1990inertial, strange2016dirac}.
Through systematic non-relativistic reduction, the off-diagonal components of the spacetime metric yield effective gauge potentials that encode the mechanical rotation \cite{matsuo2011spin, heifetz2022inertio}.
Specifically, in the rotating-frame Hamiltonian, the Coriolis term algebraically plays the role of a magnetic coupling, reflecting the same minimal-coupling substitution that is fundamental to electrodynamics.
This correspondence leads to observable macroscopic spin-rotation couplings \cite{rizzi2003sagnac, ono2015barnett, matsuo2011effects}.

Topological geometric phases and Sagnac shifts are observable consequences of these emergent gauge fields in optical interferometry \cite{tomita1986observation, stedman1997ring}, subsequently generalized to arbitrary non-circular paths \cite{wang2004generalized}.
Beyond optical interferometry, the Coriolis--Lorentz analogy implies that rotational effects extend naturally to quantum many-body systems.
In bulk quantum gases, rapid rotation restructures the system topology by driving Bose--Einstein condensates into the lowest Landau level \cite{schweikhard2004rapidly, fetter2009rotating}.
At the mesoscopic scale, this rotational physics finds concrete expression in solid-state systems, where persistent currents can be induced in quantum rings even in the absence of external magnetic fields \cite{vignale1995persistent, zivkovic2008sagnac}.
Exact treatments of rotation-modified quantum rings have recently been reported \cite{pereira2025exact}.
Coherently coupling the resulting Sagnac phases with Aharonov--Bohm fluxes provides a robust mechanism for designing ultrasensitive nanoscale interferometers \cite{search2009sagnac}.
Recent Sagnac-like measurements in atomic waveguides \cite{moan2020quantum} and gapless Dirac materials \cite{fesh2024sagnac} further attest to the universality of the Sagnac phase across modern quantum platforms.

These advances underscore a unifying thread: rotation effectively acts as a tunable gauge field.
Equally profound, however, is the role of intrinsic geometric confinement, which independently generates effective potentials through the curvature of the constraint manifold itself.
The emerging field of curvilinear electronics has established that nanoscale geometry can modulate charge localization and resistivity \cite{gentile2022electronic, ono2009tuning}.
The precise description of quantum particles confined to curved manifolds was clarified by the thin-layer quantization procedure \cite{jensen1971quantum, dacosta1981quantum, dacosta1982constraints}.
The formalism shows that tight transverse confinement inevitably gives rise to a purely quantum geometric potential, the well-known da Costa potential.
It depends quadratically on the mean curvature and linearly on the Gaussian curvature of the embedding manifold \cite{maraner1995complete, schuster2003quantum, stengel2018quantum}.

Subsequent studies established that mechanical twisting and prescribed strain gradients can strongly amplify the geometric potential and generate non-trivial geometric phases in the wavefunction \cite{ortix2011curvature, dasilva2017quantum, delgado2026NonRelativistic}.
For one-dimensional (1D) space curves, the conventional Frenet--Serret frame introduces a pathological mixing of the transverse momentum into the tangential dynamics.
The resulting frame dependence is removed by abandoning the Frenet--Serret frame in favor of the rotation-minimizing Bishop frame, whose parallel-transported transverse basis eliminates the torsion-induced coordinate singularity \cite{dasilva2017differential}. For the nondegenerate transverse ground state, however, the torsion does not enter the local effective Hamiltonian at all \cite{hou2026torsion}; the frame choice matters because it makes this absence manifest and keeps the transverse basis free of the Frenet--Serret singularity, while the coupling that does survive the frame is removed by the parity of the transverse well rather than by the frame.

When geometric confinement, external electromagnetic fields, and non-inertial rotation are simultaneously considered, the gauge coupling becomes far more involved.
In particular, the interplay among these ingredients gives rise to non-trivial adiabatic geometric phases \cite{berry1984quantal, mitchell2001gauge, atanasov2015curvature}.
For rotating two-dimensional (2D) closed surfaces, the combination of rotation and an external magnetic field can significantly modify Landau-level degeneracies and produce complex topological phenomena \cite{lima2018effects, schmidt2020exact, wang2022geometry}.
Moreover, a direct dimensional reduction in magnetic fields gives rise to an apparent transverse-field-curvature coupling \cite{encinosa2006coupling}---a gauge-dependent artifact subsequently eliminated through a suitable transverse gauge \cite{ferrari2008schrodinger}.
This gauge choice can be unified under the broader framework of a generalized transverse Poincar\'{e} gauge, which eliminates transverse field components without relying on ad hoc prescriptions.
For 1D space curves, the situation is considerably more delicate.
Even within the optimal Bishop frame, transverse external fields  couple to the tangential dynamics as first-order linear perturbations that no choice of frame can remove; only a symmetry of the transverse well can \cite{ortix2015quantum, ortix2010effect, bulaev2008spin}.
The shape of the transverse cross-section also enters the effective 1D dynamics of a space curve \cite{wang2018geometric}.
This is not a purely formal concern: in helical nanowires and spirally rolled-up nanotubes, the components of the field normal to the wire are not small.
Recent work explicitly cautions that neglecting these transverse degrees of freedom renders conventional 1D models inadequate for capturing correct topological spin textures \cite{peng2019spinorbit}.
This residual cross-coupling obstructs the precise adiabatic separation of variables, necessitating a new decoupling mechanism to restore a pure 1D tangential dynamics.

In this paper, we develop a gauge-consistent framework for the quantum dynamics of a charged particle confined to a rotating curved manifold in the presence of external electromagnetic fields, in which the tangential and transverse dynamics decouple to first order in the transverse field.
We resolve three interdependent issues that obstruct a clean separation of tangential and transverse dynamics: (i) the spurious transverse-field-curvature coupling of 2D dimensional reduction, (ii) the torsion-induced frame rotation of the conventional Frenet--Serret description of 1D curves, and (iii) the first-order transverse magnetic coupling that survives both the gauge choice and the optimal frame. We refer to this obstruction collectively as gauge--geometry entanglement: the gauge removes (i) at zeroth order, the frame removes (ii) identically, and the parity of the transverse ground state removes (iii) in both geometries.
Once these couplings are removed, an axial field leaves a joint Sagnac--Aharonov--Bohm residual of tunable sign on the torus without spoiling the poloidal--azimuthal separation, whereas an in-plane field is annihilated to first order on a planar 1D curve by the parity selection rule. Relative to the static-surface treatment of Ref.~\cite{ferrari2008schrodinger}, the present framework extends the same gauge construction to rotating frames and to 1D curves, and replaces the assumption of a weak transverse field by the parity selection rule.
The paper is organized as follows.
In Sec.\ \ref{sec:three-dimensional}, we perform the non-relativistic reduction directly from the covariant KG equation, incorporating all inertial effects while preserving the gauge symmetry.
In Sec.\ \ref{sec:two-dimensional}, for 2D surfaces, we introduce a generalized transverse Poincar\'{e} gauge that systematically absorbs macroscopic magnetic effects into the tangential vector potential, eliminating spurious curvature-field couplings, and identify the parity selection rule that removes the residual first-order field-gradient coupling.
And then we apply this formalism to a rotating toroidal surface in a static magnetic field, where the geometric potential localizes the amplitude at the innermost edge, the axial field tunes this well through a joint Sagnac--Aharonov--Bohm residual, and the in-plane field breaks the axial symmetry.
In Sec.\ \ref{sec:one-dimensional}, for space curves in the Bishop frame, we demonstrate that the deep confining potential enforces the same parity selection rule as in the 2D case; adiabatic tracing out of the transverse mode exactly annihilates all odd-parity first-order perturbations, restoring a complete separation of variables.
We further investigate a corrugated quantum ring, in which periodic geometric and flux modulations jointly generate artificial superlattices up to $\mathcal{O}(\epsilon^2)$.
Its planar geometry additionally renders the 1D tangential dynamics insensitive to the in-plane field to first order.
Finally, in Sec.~\ref{sec:conclusion}, we summarize our results and discuss their implications.

\section{Three-dimensional quantum mechanics in a rotating frame}\label{sec:three-dimensional}

We consider a spinless particle of mass $m$ and charge $Q$ minimally coupled to an external electromagnetic field.
The dynamics of the scalar field $\phi$ is governed by the covariant KG equation:
\begin{equation}
    \label{eq:kg}
    \frac{1}{\sqrt{-\eta}} D_\mu \left( \sqrt{-\eta} \,\eta^{\mu\nu} D_\nu \phi \right) + \frac{m^2c^2}{\hbar^2}\phi = 0,
\end{equation}
where $\eta$ is the determinant of the metric $\eta_{\mu\nu}$, the minimal coupling is encapsulated in the gauge-covariant derivative $D_\mu = \partial_\mu + \mathrm{i}Q/(\hbar c)A_\mu$, with $A_\mu$ the covariant electromagnetic four-potential.
Greek indices $\mu,\nu = 0,1,2,3$ are spacetime indices.

Let the reference frame $x^\mu = (ct, \bm{r})$ rotate uniformly with angular velocity $\bm{\omega}$.
The local velocity is $\bm{v} = \bm{\omega} \times \bm{r}$.
The spacetime metric $\eta_{\mu\nu}$ and its contravariant inverse $\eta^{\mu\nu}$, the latter following from the Schur complement for block matrices, take the form
\begin{equation}
    \label{eq:metric}
\begin{aligned}
    \eta_{00} &= 1 - \frac{\bm{v}^2}{c^2}, \quad \eta_{0i} = -\frac{v_i}{c}, \quad \eta_{ij} = -\delta_{ij}, \\
    \eta^{00} &= 1, \quad \eta^{0i} = -\frac{v_i}{c}, \quad \eta^{ij} = -\delta_{ij} + \frac{v_i v_j}{c^2},
\end{aligned}
\end{equation}
with spatial indices $i,j = 1,2,3$ referring to the Cartesian coordinates, and the metric determinant satisfies $\sqrt{-\eta} = 1$.
In this local coordinate system, the covariant electromagnetic four-potential is parameterized by the scalar and vector potentials as $A_\mu = (\varphi, -\bm{A})$.
Substituting the metric, Eq.~\eqref{eq:metric}, and the decomposed potentials into the covariant KG equation, Eq.~\eqref{eq:kg}, we obtain
\begin{equation}
  \label{eq:kg-rotating}
  \left[ \left(cD_0 - \bm{v} \cdot \bm{D}\right)^2 - c^2 \bm{D}^2 + \frac{m^2c^4}{\hbar^2} \right] \phi(\bm{r},t) = 0,
\end{equation}
where $\bm{D} = \nabla - \mathrm{i}Q/(\hbar c)\bm{A}$ is the three-dimensional (3D) gauge-covariant derivative and $cD_0 = \partial_t + \mathrm{i}Q\varphi/\hbar$, with $D_0$ the time component of $D_\mu$.
The cross-coupling carried by the off-diagonal metric components $\eta^{0i}$ is contained in the combination $cD_0 - \bm{v} \cdot \bm{D}$, which couples the temporal evolution to spatial gradients; in the non-relativistic limit this combination reduces to the Coriolis term $\bm{\omega}\cdot\hat{\bm{L}}$ and the associated Sagnac phase \cite{mashhoon1988neutron, hehl1990inertial, vignale1995persistent}.

In the low-energy regime, we adopt the standard non-relativistic ansatz $\phi(\bm{r},t) = \exp(-\mathrm{i}mc^2t/\hbar) \psi(\bm{r},t)$.
Under the slowly varying envelope approximation $|\partial_t \psi| \ll (mc^2/\hbar)|\psi|$ and the low-speed limit $|\bm{v}| \ll c$, the second-order time derivative and $\mathcal{O}(|\bm{v}|^2/c^2)$ corrections vanish, canceling the rest-mass term.
This reduction yields the effective 3D Schr\"{o}dinger equation:
\begin{equation}
  \begin{multlined}
    \mathrm{i}\hbar\,\partial_t \psi = \left[ \frac{1}{2m}\left(\hat{\bm{p}} - \frac{Q}{c}\bm{A}\right)^2 + Q\varphi
    - \bm{\omega}\cdot\hat{\bm{L}} \right.\\
    \left. + \frac{Q}{c}\bm{\omega}\cdot(\bm{r}\times\bm{A}) \right] \psi,
  \end{multlined}
\end{equation}
where $\hat{\bm{p}} = -\mathrm{i}\hbar\,\nabla$ and $\hat{\bm{L}} = \bm{r} \times \hat{\bm{p}}$ are the canonical momentum and orbital angular momentum operators.

The Coriolis term and the magnetic-rotational coupling are two facets of a single covariant structure: they combine into the coupling $-\bm{\omega}\cdot[\bm{r}\times(\hat{\bm{p}} - (Q/c)\bm{A})]$ to the mechanical angular momentum.
Because this coupling is gauge invariant, the reorganization below is an algebraic rewriting rather than a choice of gauge.
Completing the square for the Coriolis term and the magnetic-rotational coupling absorbs all inertial effects into effective electromagnetic potentials and recasts the Hamiltonian in the manifestly gauge-covariant form:
\begin{equation}
    \label{eq:eff-hamiltonian}
    \mathrm{i}\hbar\,\partial_t \psi = \left[ \frac{1}{2m}\left(\hat{\bm{p}} - \frac{Q}{c}\tilde{\bm{A}}\right)^2 + Q\tilde{\varphi} \right] \psi,
\end{equation}
where the effective vector potential and the effective scalar potential are defined respectively as
\begin{equation}
    \label{eq:eff-potentials}
    \tilde{\bm{A}} = \bm{A} + \frac{mc}{Q}(\bm{\omega}\times\bm{r}), \quad \tilde{\varphi} = \varphi - \frac{m}{2Q}(\bm{\omega}\times\bm{r})^2.
\end{equation}
This establishes an equivalence between the non-inertial fictitious forces and the classical electromagnetic fields, up to corrections of $\mathcal{O}(|\bm{v}|^2/c^2)$.
Specifically, the effective magnetic field
\begin{equation}
    \tilde{\bm{B}} = \nabla \times \tilde{\bm{A}} = \bm{B} + \frac{2mc}{Q}\bm{\omega}
\end{equation}
acts as a uniform axial magnetic field, recovering the quantum-mechanical formulation of Larmor's theorem \cite{vignale1995persistent}.
Similarly, the effective electric field
\begin{equation}
    \tilde{\bm{E}} = -\nabla\tilde{\varphi} - c^{-1}\partial_t\tilde{\bm{A}} = \bm{E} + \frac{m}{Q}\omega^2\bm{r}_\perp
\end{equation}
reproduces the classical centrifugal force.
Here, $\bm{r}_\perp = \bm{r} - (\bm{r}\cdot\hat{\bm{\omega}})\hat{\bm{\omega}}$ is the transverse radial vector perpendicular to the rotation axis, with $\hat{\bm{\omega}}$ being the unit vector along the rotation axis, $\bm{\omega} = \omega\,\hat{\bm{\omega}}$.
The mapping to a static inertial frame thus preserves the $U(1)$ gauge structure of the original problem.
Larmor's theorem thus extends from uniform rotation to the constrained manifolds treated below: results established for magnetic mesoscopic systems carry over to rotating ones under the same substitution of potentials.

The applications below use fields that are static in the laboratory frame. Expressed in the rotating frame, such a field carries, besides the rotated vector potential, the scalar component $\varphi=-(\bm v\cdot\bm A)/c$, whose gradient reproduces the Lorentz force $(\bm v/c)\times\bm B$. Two configurations are therefore possible. (A) For a pure static magnetic field the rotating-frame scalar potential is the motional one, $\varphi=-(\bm v\cdot\bm A)/c$; the cross term $(Q/c)\bm\omega\cdot(\bm r\times\bm A)$ is then absorbed into $Q\varphi$, and the Hamiltonian reduces to its standard rotating-frame form. (B) If the laboratory field also contains the static electrostatic potential $\varphi_\mathrm{lab}=(\bm v\cdot\bm A_\mathrm{lab})/c$, which is time independent because neither $\bm v$ nor $\bm A_\mathrm{lab}$ depends explicitly on time, the rotating-frame scalar potential vanishes, $\varphi=0$. The applications in Secs.~\ref{sec:two-dimensional} and \ref{sec:one-dimensional} refer to configuration (B); configuration (A) follows from the general equations by the substitution $\varphi\to-(\bm v\cdot\bm A)/c$, which removes the joint Sagnac--Aharonov--Bohm term from the effective potential and leaves it in the azimuthal kinetic operator, where it shifts the spectrum and the persistent current.

Equipped with the effective potentials of Eq.~\eqref{eq:eff-potentials}, we next recast Eq.~\eqref{eq:eff-hamiltonian} in a general 3D curvilinear coordinate system $(q^1, q^2, q^3)$, in preparation for constraining the particle to a curved low-dimensional manifold:
\begin{equation}
    \mathrm{i}\hbar \tilde{D}_0 \psi = \left[ -\frac{\hbar^2}{2m} \frac{1}{\sqrt{G}} \tilde{D}_i \left( \sqrt{G} \, G^{ij} \tilde{D}_j \right) + V_\mathrm{c} \right] \psi.
    \label{eq:3d-curvilinear-schrodinger}
\end{equation}
Here and in the following, the Latin indices $i, j = 1, 2, 3$ stand for the 3D curvilinear coordinates, $G^{ij}$ is the contravariant metric tensor, and $G = \det(G_{ij})$.
A steep confining potential $V_\mathrm{c}$, co-rotating with the structure, is assumed to localize the particle on the target constraint manifold, providing the basis for the subsequent thin-layer quantization.
Beyond steepness, the well is required to be a single one, so that its transverse ground state is non-degenerate, and its transverse level spacing $\Delta E_\perp \equiv E_1 - E_0$ is required to exceed every other energy scale of the problem---the tangential kinetic energy $\hbar^2/(2mL^2)$ for a manifold of size $L$, the rotational scale $\hbar\omega$, and the magnetic scale $\hbar|Q\tilde{B}|/(mc)$.
A well with a (nearly) degenerate transverse ground state---a symmetric double well, for instance, whose $\Delta E_\perp$ is set by tunnelling and may be exponentially small while its transverse extent remains large---falls outside this assumption and is not considered here (see Appendix~\ref{app:bounds}).
The operators $\tilde{D}_0$ and $\tilde{D}_i$ represent the effective temporal and spatial gauge-covariant derivatives, inherently incorporating the rotation-modified scalar and vector potentials:
\begin{equation}
    \label{eq:covariant-derivatives}
    \tilde{D}_0 = \partial_t + \frac{\mathrm{i}Q}{\hbar}\tilde{\varphi}, \quad \tilde{D}_i = \partial_i - \frac{\mathrm{i}Q}{\hbar c}\tilde{A}_i,
\end{equation}
where $\partial_i \equiv \partial/\partial q^i$.
Constraining the particle to a lower-dimensional manifold couples the tangential and transverse dynamics through the induced geometric and gauge-field contributions.
The next two sections apply the thin-layer quantization to Eq.~\eqref{eq:3d-curvilinear-schrodinger} and remove these obstructions for a 2D curved surface and a 1D space curve, respectively.

\section{Quantum mechanics on a two-dimensional curved surface}\label{sec:two-dimensional}

\subsection{General formalism and transverse Poincar\'{e} gauge}\label{subsec:2d-formalism}

\begin{figure}[!ht]
  \centering
  \begin{minipage}[t]{0.48\columnwidth}
    \centering
    \begin{overpic}[width=\linewidth]{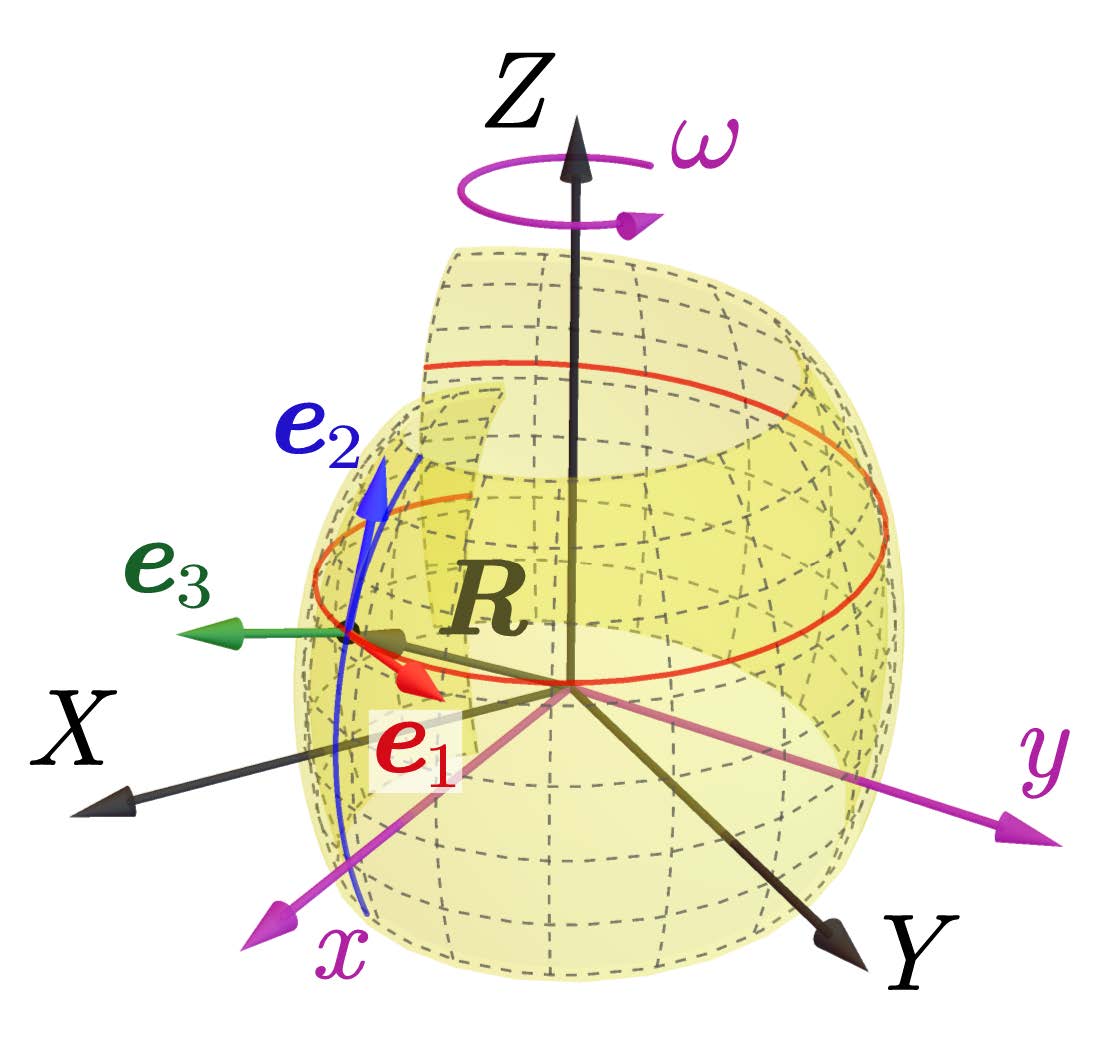}
      \put(0,85){$(a)$}
    \end{overpic}
  \end{minipage}
  \hfill
  \begin{minipage}[t]{0.48\columnwidth}
    \centering
    \begin{overpic}[width=\linewidth]{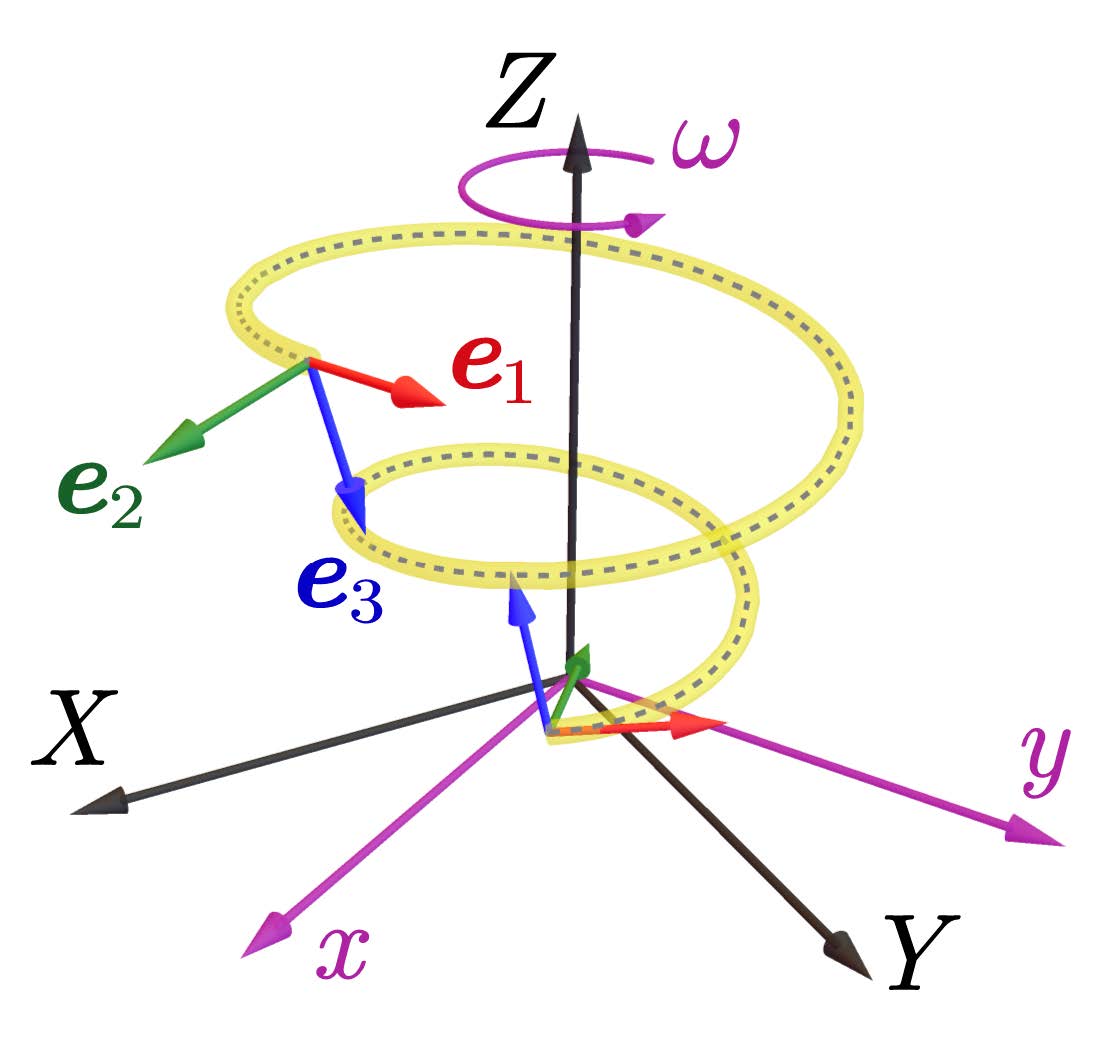}
      \put(0,85){$(b)$}
    \end{overpic}
  \end{minipage}
  \caption{%
    \label{fig:schematic}
    Constrained low-dimensional quantum systems in a uniformly rotating frame.
    The non-inertial frame $(x, y, z)$ rotates with an arbitrary constant angular velocity $\bm{\omega}$ relative to the inertial frame $(X, Y, Z)$.
    $(a)$ 2D curved surface, with the adapted curvilinear basis: tangent vectors $\bm{e}_1, \bm{e}_2$ and unit normal vector $\bm{e}_3$.
    $(b)$ 1D space curve, with the rotation-minimizing Bishop frame: unit tangent vector $\bm{e}_1$ and parallel-transported transverse basis vectors $\{\bm{e}_2, \bm{e}_3\}$.
  }
\end{figure}

Following the thin-layer procedure of da Costa \cite{dacosta1981quantum}, we parameterize the spatial neighborhood of a 2D surface $S$ by the adapted curvilinear coordinate system shown in Fig.\ \ref{fig:schematic} $(a)$.
Let $\bm{r}(q^1, q^2)$ be the position vector of $S$, where $q^a$ ($a = 1, 2$) denote intrinsic surface coordinates.
And then the 3D space surrounding $S$ can be parameterized as
\begin{equation}
  \bm{R}(q^1, q^2, q^3) = \bm{r}(q^1, q^2) + q^3 \bm{e}_3(q^1, q^2),
\end{equation}
where $\bm{e}_3$ is the unit basis vector normal to $S$, and $|q^3|$ measures the distance from $S$ along $\bm{e}_3$.
The covariant 3D metric tensor is defined as $G_{ij} = \partial_i \bm{R} \cdot \partial_j \bm{R}$.
Normalization and orthogonality of $\bm{e}_3$ give $G_{33} = 1$ and $G_{a3} = 0$.
The surface metric components $G_{ab}$ are related to the induced 2D metric $g_{ab} = \partial_a \bm{r} \cdot \partial_b \bm{r}$ via the Weingarten equations $\partial_a \bm{e}_3 = w^b_a \partial_b \bm{r}$, yielding
\begin{equation}
    G_{ab} = g_{ab} + \left(\hat{w}\hat{g} + \hat{g}^T \hat{w}^T\right)_{ab} q^3 + \left(\hat{w}\hat{g}\hat{w}^T\right)_{ab} (q^3)^2,
\end{equation}
where $w^b_a$ are the components of the Weingarten curvature matrix $\hat{w}$.
Consequently, the metric determinants satisfy $G = f^2 g$, where the geometric scaling factor
\begin{equation}
    f = 1 + 2M q^3 + K (q^3)^2,
\end{equation}
with $M = \mathrm{Tr}(\hat{w})/2$ the mean curvature and $K = \det(\hat{w})$ the Gaussian curvature.

The kinetic term in Eq.~\eqref{eq:3d-curvilinear-schrodinger} is governed by the gauge-covariant momentum
\begin{equation}
    \hat{\Pi}_i = -\mathrm{i}\hbar\,\partial_i - \frac{Q}{c}\tilde{A}_i,
\end{equation}
where $\partial_i \equiv \partial/\partial q^i$ and $\tilde{A}_i$ are the covariant components in the da Costa curvilinear basis.
An adiabatic separation of the tangential and transverse degrees of freedom faces two obstructions.
First, the Laplace--Beltrami operator acquires a first-order normal derivative $\partial_3$, which obstructs any direct factorization of the kinetic energy into independent tangential and transverse parts.
Second, the transverse component $\tilde{A}_3$ of the effective vector potential couples the tangential and transverse dynamics through the gauge-covariant kinetic term.
The former is eliminated by introducing a new wave function $\chi = f^{1/2}\psi$, which absorbs the Jacobian of the coordinate transformation.
The latter is eliminated by the generalized transverse Poincar\'{e} gauge, which, taking the curved surface $S$ ($q^3 = 0$) as the reference manifold, defines the scalar gauge function by integrating $\tilde{A}_3$ along the ray normal to $S$, following the Poincar\'{e} (multipolar) line-integral construction \cite{brittin1982poincare,ellis1990dirac} but with the reference point generalized from a point to the manifold $S$:
\begin{equation}
    \label{eq:poincare-2d}
    \gamma(q^1, q^2, q^3) = -\int_0^1 q^3 \tilde{A}_3(q^1, q^2, \lambda q^3) \,\mathrm{d}\lambda,
\end{equation}
where $\lambda \in [0,1]$ is a dimensionless integration parameter.
The gauge transformation $\tilde{\bm{A}} \to \tilde{\bm{A}}' = \tilde{\bm{A}} + \nabla\gamma$ then identically cancels the transverse component: $\tilde{A}'_3 = 0$.
Under the change of variables $z = \lambda q^3$, this gauge function reduces to the transverse gauge employed by Ferrari and Cuoghi for static surfaces \cite{ferrari2008schrodinger}.
All electromagnetic effects are thereby confined to the tangential components.

With the transverse gauge coupling eliminated and the strong transverse confinement enforcing an adiabatic separation, the rescaled wavefunction factorizes as $\chi(q^1, q^2, q^3, t) = \chi_\mathrm{t}(q^1, q^2, t) \chi_\mathrm{n}(q^3, t)$, where $\chi_\mathrm{t}$ and $\chi_\mathrm{n}$ denote the tangential and transverse components, respectively.
The transverse component satisfies a 1D Schr\"{o}dinger equation
\begin{equation}
    \mathrm{i}\hbar\,\partial_t \chi_\mathrm{n}(q^3, t) = \left[ -\frac{\hbar^2}{2m}\partial_3^2 + V_\mathrm{c}(q^3) \right] \chi_\mathrm{n}(q^3, t),
\end{equation}
whose ground state $\chi^{(0)}_\mathrm{n}(q^3)$ is determined solely by the confining potential $V_\mathrm{c}(q^3)$.
Owing to the steepness of $V_\mathrm{c}$, the transverse level spacing greatly exceeds every other energy scale in the problem, so that the particle is frozen in $\chi^{(0)}_\mathrm{n}$ and the associated dynamical phase factor $\exp(-\mathrm{i}E^{(0)}_\mathrm{n}t/\hbar)$ is integrated out.

The tangential wavefunction $\chi_\mathrm{t}$ then obeys the effective 2D Schr\"{o}dinger equation:
\begin{equation}
    \label{eq:2d-effective}
    \mathrm{i}\hbar \tilde{D}_0 \chi_\mathrm{t} = \left[ \frac{1}{2m} \frac{1}{\sqrt{g}} \hat{\Pi}_a \left( \sqrt{g} \, g^{ab} \, \hat{\Pi}_b \right) + V_\mathrm{g}^\text{2D} \right] \chi_\mathrm{t},
\end{equation}
where $\hat{\Pi}_a = -\mathrm{i}\hbar\,\partial_a - (Q/c)\tilde{A}_a$ ($a = 1,2$) is the 2D gauge-covariant momentum evaluated on the reference surface $S$ ($q^3 = 0$).
On $S$, the Poincar\'{e} gauge leaves the tangential components unchanged, since $\partial_a\gamma|_{q^3=0}=0$, whence $\tilde{A}'_a = \tilde{A}_a$.
The da Costa geometric potential
\begin{equation}
    \label{eq:dacosta-2d}
    V_\mathrm{g}^\text{2D} = -\frac{\hbar^2}{2m}(M^2 - K)
\end{equation}
emerges from the transverse ground-state expectation of the $q^3$-dependent metric terms.
Constructed from the metric alone, $V_\mathrm{g}^\text{2D}$ is independent of both the external field and the rotation.
In the thin-layer limit, the rotating 2D system is therefore governed solely by the tangential vector potential and the geometric potential $V_\mathrm{g}^\text{2D}$, the transverse field components being completely decoupled.

\subsection{Validity limits and symmetry considerations}\label{subsec:2d-validity}

Eq.~\eqref{eq:2d-effective} is exact on the reference surface $S$, but a real layer has a finite width, over which the tangential components of the vector potential vary.
Although the Poincar\'{e} gauge has eliminated $\tilde{A}_3$ on the surface, retaining the first-order variation of the gauge-transformed tangential components away from $S$ yields, to first order in $q^3$, the residual coupling
\begin{equation}
\begin{multlined}
    \tilde{A}'_a(q^1, q^2, q^3) \approx \tilde{A}_a(q^1, q^2, 0) \\
    + q^3\bigl[\partial_3 \tilde{A}_a(q^1, q^2, 0) - \partial_a \tilde{A}_3(q^1, q^2, 0)\bigr],
\end{multlined}
\end{equation}
the bracketed coefficient being exactly the transverse magnetic field components, which therefore feed into the tangential dynamics solely through the odd-parity operator $q^3$.
Two independent effects annihilate this coupling.
First, for a confining potential with spatial parity symmetry, $V_\mathrm{c}(q^3) = V_\mathrm{c}(-q^3)$, the ground state $\chi^{(0)}_\mathrm{n}$ is an even function of $q^3$ and all odd-parity expectation values vanish identically, $\langle q^3 \rangle = \langle \partial_3 \rangle = 0$.
Note that this parity selection rule---familiar from the thin-wire context \cite{ortix2015quantum} and independent of the intrinsic geometry of $S$---annihilates the coupling to first order in the magnetic field.
Second, in the strict thin-layer limit the transverse density collapses onto $S$, so that the effective potentials are evaluated exactly at $q^3 = 0$ even without parity symmetry.

In a realistic thin film, however, $V_\mathrm{c}(q^3)$ may lose perfect parity through built-in electric fields or strain gradients, so that the transverse wavefunction acquires a finite mean position, $\langle q^3 \rangle \approx F_\perp d_\mathrm{rms}^2/\Delta E_\perp$ (see Appendix~\ref{app:bounds} for the perturbative estimate).
Here, $F_\perp$ is the transverse force, $d_\mathrm{rms} = \langle (q^3)^2 \rangle_0^{1/2}$ is the per-direction root-mean-square width of the transverse ground state, and $\Delta E_\perp$ is the transverse level spacing.
Such a displacement would generate Stark shifts and renormalize the effective curvatures, but it vanishes in the strict thin-layer limit.
It remains perturbative as long as the induced energy shift stays negligible, $F_\perp d_\mathrm{rms} \ll \Delta E_\perp$.
The same estimate covers the normal variation of the effective scalar potential $\tilde{\varphi}$, which enters Eq.~\eqref{eq:2d-effective} only through its surface value (Appendix~\ref{app:bounds}).
Consequently, Eq.~\eqref{eq:2d-effective} remains accurate as long as the tangential vector potential varies little across the layer,
\begin{equation}
    d_\mathrm{rms}\,\bigl|\partial_3 \tilde{A}_a - \partial_a \tilde{A}_3\bigr| \ll \bigl|\tilde{A}_a\bigr| .
\end{equation}

\subsection{Application to a rotating torus in a static magnetic field}\label{subsec:2d-torus}

\begin{figure}[!tbp]
  \centering
  \begin{minipage}[t]{0.48\columnwidth}
    \vspace{0pt}
    \centering
    \begin{overpic}[width=\textwidth]{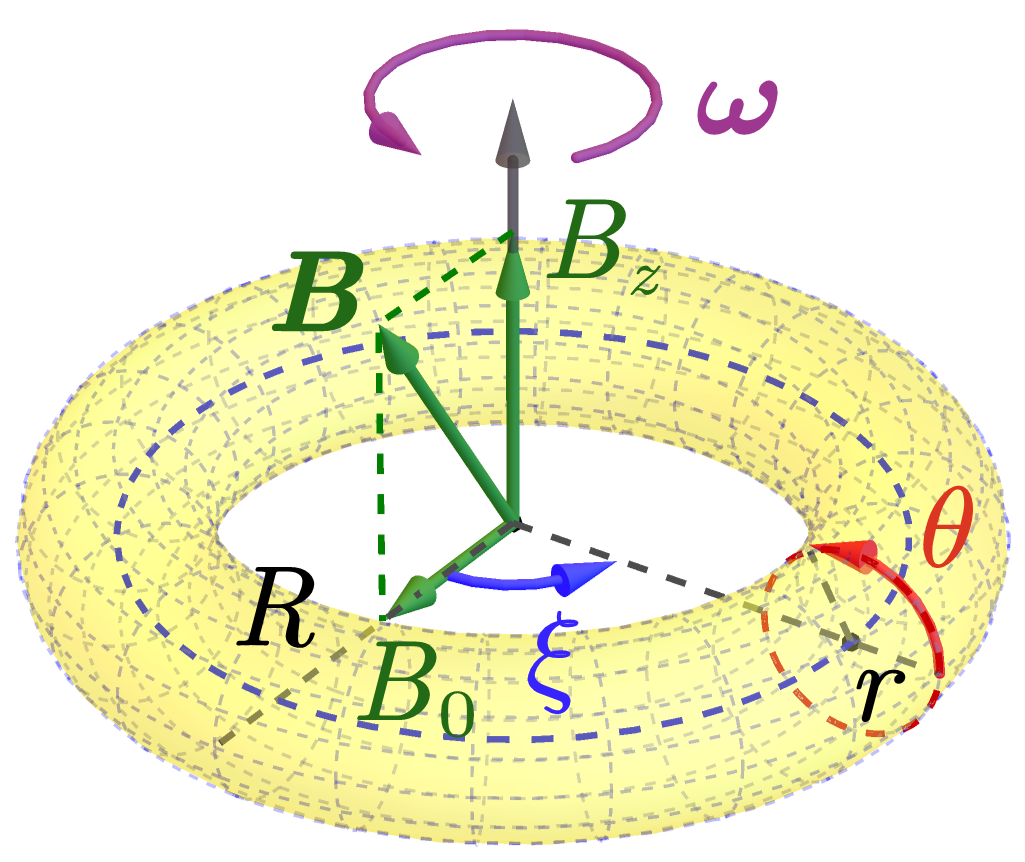}
      \put(0,78){$(a)$}
    \end{overpic}
  \end{minipage}%
  \hfill
  \begin{minipage}[t]{0.48\columnwidth}
    \vspace{0pt}
    \centering
    \begin{overpic}[width=\textwidth]{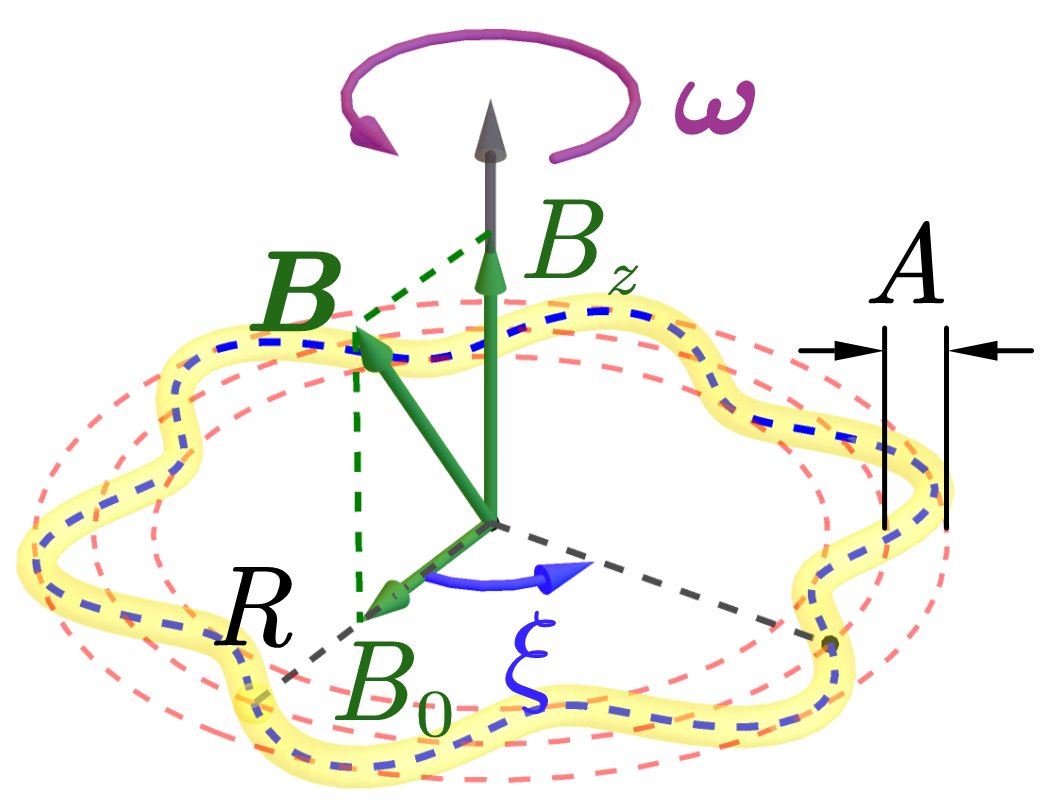}
      \put(0,70){$(b)$}
    \end{overpic}
  \end{minipage}
  \caption{%
    \label{fig:applications}
    Low-dimensional curved structures rotating with a constant angular velocity $\bm{\omega}$ in a laboratory-static magnetic field $\bm{B} = B_0\hat{\bm{x}} + B_z\hat{\bm{z}}$, with $B_0$ the in-plane component, perpendicular to the rotation axis, and $B_z$ the axial component.
    $(a)$ A 2D toroidal surface characterized by its major radius $R$ and minor radius $r$, parameterized by the poloidal angle $\theta$ and the azimuthal angle $\xi$.
    The curvature-induced da Costa potential depends solely on $\theta$, whereas the rotation enters the azimuthal momentum through the Sagnac vector potential.
    $(b)$ A 1D corrugated quantum ring characterized by its base radius $R$ and a periodic radial modulation of amplitude $A$, parameterized by the azimuthal angle $\xi$.
    The periodic curvature modulation and the effective centrifugal potential together generate an artificial superlattice.
  }
\end{figure}

To concretely illustrate the spatial decoupling mechanism, we apply the formalism to a charged particle on a toroidal surface rotating uniformly about its symmetry axis with $\bm{\omega} = \omega\,\hat{\bm{z}}$.
The torus is immersed in a laboratory-static magnetic field $\bm{B} = B_0\hat{\bm{x}} + B_z\hat{\bm{z}}$, with $B_z$ the axial component, along the rotation axis, and $B_0$ the in-plane component, lying in the plane perpendicular to that axis.
Since the field is static in the laboratory frame while the torus is at rest in the rotating frame, the in-plane component is seen to rotate about the rotation axis at frequency $\omega$, whereas the axial component remains constant; this rotation is the origin of the explicit time dependence derived below.
As depicted in Fig.\ \ref{fig:applications} $(a)$, the torus has major radius $R$, the distance from the symmetry axis to the centre of the tube, and minor radius $r$, the radius of the tube's cross-section, with $R > r$; it is parameterized by the poloidal angle $\theta \in [0, 2\pi)$ and the azimuthal angle $\xi \in [0, 2\pi)$.
Introducing the local radial distance from the symmetry axis, $\rho(\theta) = R + r\cos\theta$, its position vector and induced metric reduce to $\bm{R}(\theta, \xi) = (\rho(\theta)\cos\xi,\, \rho(\theta)\sin\xi,\, r\sin\theta)$ with $g_{\theta\theta} = r^2$, $g_{\xi\xi} = \rho^2(\theta)$, and $g_{\theta\xi} = 0$.
The principal curvatures $k_\theta = 1/r$ and $k_\xi = \cos\theta/\rho(\theta)$, substituted into Eq.~\eqref{eq:dacosta-2d}, yield the exact da Costa potential
\begin{equation}
    V_\mathrm{g}^\text{2D}(\theta) = -\frac{\hbar^2}{8m}\left( \frac{1}{r} - \frac{\cos\theta}{\rho(\theta)} \right)^2.
\end{equation}
This potential forms an attractive geometric well along the poloidal direction, localizing the amplitude at the innermost edge $\theta = \pi$.
The rotational terms of Eq.~\eqref{eq:eff-potentials} then give, for $\bm{\omega} = \omega\hat{\bm{z}}$, the covariant components $\tilde{A}_{\mathrm{rot},\theta} = 0$ and $\tilde{A}_{\mathrm{rot},\xi}(\theta) = (mc\omega/Q)\rho^2(\theta)$, together with the centrifugal scalar potential $V_\mathrm{cf}(\theta) = -m\omega^2\rho^2(\theta)/2$.
Expressing the laboratory field in the rotating frame and adopting the symmetric gauge, the covariant components of the external potential are exactly
\begin{equation}
\begin{split}
    \tilde{A}^{\text{ext}}_\theta &= \frac{1}{2}B_0\, r\,(r + R\cos\theta)\,\sin(\xi + \omega t), \\
    \tilde{A}^{\text{ext}}_\xi &= \frac{1}{2}B_z\rho^2(\theta) - \frac{1}{2}B_0\, r\,\rho(\theta)\sin\theta\,\cos(\xi + \omega t),
\end{split}
\end{equation}
so that, since $\tilde{A}_{\mathrm{rot},\theta} = 0$, the poloidal component is unchanged, $\tilde{A}_\theta = \tilde{A}^{\text{ext}}_\theta$, and only the azimuthal one becomes
\begin{equation}
    \tilde{A}_\xi(\theta, \xi, t) = \frac{1}{2}B_\mathrm{eff}\rho^2(\theta) - \frac{1}{2}B_0\, r\,\rho(\theta)\sin\theta\,\cos(\xi + \omega t),
\end{equation}
with $B_\mathrm{eff} = B_z + 2mc\omega/Q$, which is the Larmor mapping of the axial field onto the rotation. In configuration (B) of Sec.~\ref{sec:three-dimensional} the rotating-frame scalar potential vanishes, and Eq.~\eqref{eq:residual-Bz} then follows from this expression.
According to Eq.~\eqref{eq:2d-effective}, the tangential wavefunction $\chi_\mathrm{t}(\theta, \xi, t)$ satisfies
\begin{multline}
    \label{eq:torus-hamiltonian}
    \mathrm{i}\hbar\,\partial_t \chi_\mathrm{t} = \Bigg[ \frac{1}{2m}\left( \frac{1}{r^2\rho(\theta)}\hat{\Pi}_\theta\left( \rho(\theta)\hat{\Pi}_\theta \right) + \frac{\hat{\Pi}_\xi^2}{\rho^2(\theta)} \right) \\
    + V_\mathrm{g}^\text{2D}(\theta) + V_\mathrm{cf}(\theta) \Bigg] \chi_\mathrm{t},
\end{multline}
where $\hat{\Pi}_\theta = -\mathrm{i}\hbar\,\partial_\theta - (Q/c)\tilde{A}_\theta$ and $\hat{\Pi}_\xi = -\mathrm{i}\hbar\,\partial_\xi - (Q/c)\tilde{A}_\xi$.

Three regimes follow from Eq.~\eqref{eq:torus-hamiltonian}, which shows that the two components of the field enter through different operators: $B_z$ couples only through $\hat{\Pi}_\xi$, whereas $B_0$ additionally generates a poloidal component $\tilde{A}_\theta$ and thereby couples the two angles. We therefore discuss in turn the field-free rotating case, the purely axial case, and the case with an in-plane component.

For $B_0 = B_z = 0$, the azimuthal operator reduces to $\hat{\Pi}_\xi = -\mathrm{i}\hbar\,\partial_\xi - m\omega\rho^2(\theta)$ and the poloidal operator to $\hat{\Pi}_\theta = -\mathrm{i}\hbar\,\partial_\theta$, so that the diamagnetic contribution inherent in $\hat{\Pi}_\xi^2$ exactly cancels $V_\mathrm{cf}(\theta)$, the two quadratic contributions being the scalar and vector faces of one and the same rotation.
Physically, the Sagnac effect then dictates the azimuthal dynamics, whereas the intrinsic scalar geometry governs the amplitude distribution.

For $B_0 = 0$ and $B_z \neq 0$, $\tilde{A}_\theta$ still vanishes and the poloidal--azimuthal separation is preserved, but the exact cancellation is replaced by the residual
\begin{equation}
    \label{eq:residual-Bz}
    \Delta V(\theta) = \rho^2(\theta)\left[ \frac{\omega QB_z}{2c} + \frac{Q^2B_z^2}{8mc^2} \right],
\end{equation}
a joint Sagnac--Aharonov--Bohm term proportional to $B_z\left(B_z+4mc\omega/Q\right)$, whose part linear in $B_z$ carries the sign of $\omega B_z$ and is the interference of the Sagnac and Aharonov--Bohm phases, the quadratic part being positive definite; the term therefore deepens the geometric well at $\theta=\pi$ when $B_z\left(B_z+4mc\omega/Q\right)>0$ and displaces it otherwise.

By contrast, for $B_0 \neq 0$ the poloidal component $\tilde{A}_\theta$ no longer vanishes and the Hamiltonian acquires an explicit $\cos(\xi + \omega t)$ driving.
The axial symmetry is thereby broken, the poloidal and azimuthal motions are coupled, and the spatial separation is lost, the dynamics being governed by a time-periodic Hamiltonian of frequency $\omega$, i.e., a Floquet problem.
The torus thus realizes both limiting behaviours on a concrete geometry.
The axial field preserves the separation and converts the exact cancellation into a joint Sagnac--Aharonov--Bohm term of tunable sign, whereas the in-plane field destroys it through a time-periodic driving.

Two statements, one formal and one physical, close the 2D case. Formally, the generalized transverse Poincar\'{e} gauge sets $\tilde{A}'_3 = 0$ identically on $S$, so that Eq.~\eqref{eq:2d-effective} contains only the tangential potential and the da Costa potential $V_\mathrm{g}^\text{2D}$, the latter fixed by the metric alone; the coupling that survives through the finite layer thickness is odd in $q^3$ and is annihilated by the parity of the transverse ground state. Physically, the rotating torus shows that the reduced dynamics responds to the field in two qualitatively different ways: an axial field preserves the poloidal--azimuthal separation and converts the exact cancellation into a joint Sagnac--Aharonov--Bohm term of tunable sign, whereas an in-plane field destroys the axial symmetry and turns the problem into a Floquet one. The 2D obstruction is thus removed by the gauge at zeroth order in the layer thickness and by parity at first order in the field; on a 1D curve, by contrast, the further obstruction is a frame artifact rather than a gauge one: the Frenet--Serret normal rotates about the tangent, and the cure is a change of frame rather than of gauge.

\section{Quantum mechanics on a one-dimensional space curve}\label{sec:one-dimensional}

\subsection{General formalism via the Bishop frame}\label{subsec:1d-formalism}

Going from a 2D surface to a 1D curve replaces the single normal line by a normal plane whose orientation varies along the curve---an $SO(2)$ frame freedom with no 2D counterpart.
For a charged particle confined to a tubular neighborhood of a space curve $\bm{r}(q^1)$ parameterized by arc length $q^1$, the conventional Frenet--Serret frame furnishes a local orthonormal basis.
However, the geometric torsion $\tau(q^1)$ forces the normal basis to rotate around the tangent, entangling the tangential momentum with the transverse angular momentum and obstructing the adiabatic separation of the 1D dynamics.

To make the torsion-free structure manifest we replace the Frenet--Serret frame by the Bishop frame, as schematically illustrated in Fig.\ \ref{fig:schematic} $(b)$. Torsion re-enters the tangential dynamics only when a degenerate transverse band is retained \cite{hou2026torsion}; the single-band, ground-state regime considered here is the complementary one, in which the surviving obstruction is the transverse magnetic coupling rather than the torsion itself.
This frame is constructed by rotating the transverse basis through an angle $\zeta(q^1)$ satisfying
\begin{equation}
    \frac{\mathrm{d}\zeta(q^1)}{\mathrm{d}q^1} = \tau(q^1),
\end{equation}
so that the transverse basis vectors $\{\bm{e}_2, \bm{e}_3\}$ are parallel-transported---free of local rotation---along the curve.
The evolution of the Bishop frame is given by
\begin{equation}
    \partial_1 \bm{e}_1 = k_2 \bm{e}_2 + k_3 \bm{e}_3,\
    \partial_1 \bm{e}_2 = -k_2 \bm{e}_1,\
    \partial_1 \bm{e}_3 = -k_3 \bm{e}_1.
\end{equation}
Following the thin-layer procedure of Sec.~\ref{sec:two-dimensional}, the tubular neighborhood of the curve is parameterized as
\begin{equation}
    \bm{R}(q^1, q^2, q^3) = \bm{r}(q^1) + q^2 \bm{e}_2(q^1) + q^3 \bm{e}_3(q^1),
\end{equation}
where $q^2$ and $q^3$ denote the transverse local coordinates.
The metric tensor $G_{ij} = \partial_i \bm{R} \cdot \partial_j \bm{R}$ in the Bishop frame is block-diagonal: $G_{22}=G_{33}=1$, all cross-metric components $(G_{12}=G_{13}=G_{23})$ vanish, and the metric determinant reduces to $G = f^2$ with
\begin{equation}
    f(q^1, q^2, q^3) = 1 - k_2(q^1) q^2 - k_3(q^1) q^3,
\end{equation}
where $k_2(q^1)$ and $k_3(q^1)$ are the components of the curvature vector along the $\bm{e}_2$ and $\bm{e}_3$ directions, satisfying $k_2^2 + k_3^2 = k^2(q^1)$ with $k(q^1)$ the total curvature.
The transverse first-derivative terms in the Laplace--Beltrami operator are eliminated by the rescaling transformation $\chi = f^{1/2}\psi$, which absorbs the Jacobian of the coordinate transformation.
Although the spatial metric is perfectly diagonalized by the Bishop frame, the 3D effective vector potential $\tilde{\bm{A}}$ can still couple the tangential and transverse degrees of freedom via the gauge-covariant kinetic term.
To eliminate this residual coupling, we introduce a 2D generalized transverse Poincar\'{e} gauge within the transverse normal plane.
Using an auxiliary scaling parameter $\lambda \in [0, 1]$, the scalar gauge function is constructed as
\begin{equation}
    \label{eq:gauge-1d}
    \gamma(q^1, q^2, q^3) = -\int_0^1 \sum_{\alpha=2,3} q^\alpha \tilde{A}_\alpha(q^1, \lambda q^2, \lambda q^3) \,\mathrm{d}\lambda.
\end{equation}
As shown in Appendix~\ref{app:poincare-radial}, the gauge transformation $\tilde{\bm{A}} \to \tilde{\bm{A}}' = \tilde{\bm{A}} + \nabla\gamma$ removes the radial part of the transverse potential along every ray issuing from the axis, $q^2 \tilde{A}'_2 + q^3 \tilde{A}'_3 = 0$, while the azimuthal part is gauge invariant and survives.
The transverse components therefore vanish on the axis and grow only linearly away from it,
\begin{equation}
\label{eq:residual-transverse}
\begin{split}
    \tilde{A}'_2 &\simeq -\frac{1}{2} \tilde{B}_1 q^3, \
    \tilde{A}'_3 \simeq +\frac{1}{2} \tilde{B}_1 q^2,\\
    \tilde{B}_1 &\equiv \bigl(\partial_2 \tilde{A}_3 - \partial_3 \tilde{A}_2\bigr)_{q^2=q^3=0},
\end{split}
\end{equation}
which is the symmetric gauge of the tangential field $\tilde{B}_1$; substitution into the transverse kinetic energy yields the Zeeman-like and diamagnetic terms of Eq.~\eqref{eq:1d-transverse-hamiltonian}.
The first-order deviation of $\tilde{A}'_1$ away from the axis, together with the conditions under which it can be neglected, is analysed in Sec.~\ref{subsec:1d-validity}.

Owing to the adiabatic separation of energy scales in the strong-confinement limit, the rescaled wavefunction factorizes as $\chi(q^1, q^2, q^3, t) = \chi_\mathrm{t}(q^1, t) \chi_\mathrm{n}(q^2, q^3, t)$.
The transverse wavefunction $\chi_\mathrm{n}$ is governed by the transverse Hamiltonian $\hat{H}_\mathrm{n}$:
\begin{equation}
    \mathrm{i}\hbar\,\partial_t \chi_\mathrm{n} = \hat{H}_\mathrm{n} \chi_\mathrm{n},
\end{equation}
with
\begin{equation}
    \label{eq:1d-transverse-hamiltonian}
    \hat{H}_\mathrm{n} = -\frac{\hbar^2}{2m}\nabla_\mathrm{n}^2 - \frac{Q\tilde{B}_1}{2mc}\hat{L}_1 + \frac{Q^2\tilde{B}_1^2}{8mc^2}|\bm{q}_\perp|^2 + V_\mathrm{c}(q^2, q^3),
\end{equation}
where $\nabla_\mathrm{n}^2 = \partial_2^2 + \partial_3^2$.
The steep confining potential $V_\mathrm{c}(q^2, q^3)$ freezes the particle in the transverse ground state $\chi^{(0)}_\mathrm{n}(q^2, q^3)$, which is non-degenerate and is separated from the first excited transverse level by $\Delta E_\perp$ far above every other scale contained in Eq.~\eqref{eq:1d-transverse-hamiltonian} (Sec.~\ref{sec:three-dimensional}).
For a circularly symmetric $V_\mathrm{c}$, the tangential field component $\tilde{B}_1$ enters $\hat{H}_\mathrm{n}$ as a Zeeman-like coupling $\propto \tilde{B}_1 \hat{L}_1$, with $\hat{L}_1 = -\mathrm{i}\hbar\,(q^2\partial_3 - q^3\partial_2)$ denoting the orbital angular momentum about the local tangent $\bm{e}_1$.
Because the circular symmetry guarantees $[\hat{H}_\mathrm{n}, \hat{L}_1] = 0$, the transverse eigenstates carry a definite angular momentum, and the non-degenerate ground state of Eq.~\eqref{eq:1d-transverse-hamiltonian} is an $l = 0$ ($s$-wave) orbital, so that this coupling contributes nothing at first order.
The even parity of this orbital is the same property that underlies the cancellation of the transverse-field coupling in Sec.~\ref{subsec:1d-validity}; that cancellation requires only inversion symmetry, and therefore survives even for a non-circularly symmetric $V_\mathrm{c}$, for which $[\hat{H}_\mathrm{n}, \hat{L}_1] \neq 0$ but $\hat{L}_1$ is odd and its ground-state expectation still vanishes.
The diamagnetic term, proportional to the transverse cross-sectional area $\langle |\bm{q}_\perp|^2 \rangle$, vanishes in the extreme thin-wire limit.

In this limit the normal degrees of freedom are frozen in $\chi^{(0)}_\mathrm{n}$, and the tangential wavefunction $\chi_\mathrm{t}$ obeys the 1D effective Schr\"{o}dinger equation:
\begin{equation}
    \label{eq:1d-effective}
    \mathrm{i}\hbar \tilde{D}_0 \chi_\mathrm{t} = \left[ \frac{1}{2m} \hat{\Pi}_\mathrm{t}^2 + V_\mathrm{g}^\text{1D} \right] \chi_\mathrm{t},
\end{equation}
where $\tilde{D}_0 = \partial_t + \mathrm{i}Q\tilde{\varphi}/\hbar$ is the temporal gauge-covariant derivative of Eq.~\eqref{eq:covariant-derivatives}, and $\hat{\Pi}_\mathrm{t} = -\mathrm{i}\hbar\,\partial_1 - (Q/c)\tilde{A}_1$ is the 1D gauge-covariant momentum; the potentials and their derivatives are here and below evaluated on the curve axis, $q^2 = q^3 = 0$.
The 1D da Costa geometric potential,
\begin{equation}
    \label{eq:dacosta-1d}
    V_\mathrm{g}^\text{1D} = -\frac{\hbar^2 k^2(q^1)}{8m},
\end{equation}
emerges from the transverse ground-state expectation of the $q^{2,3}$-dependent metric terms.
In the extreme thin-wire limit, the rotating 1D quantum system is therefore governed solely by the tangential projection of the effective vector potential.

\subsection{Parity-induced cancellation and validity limits}\label{subsec:1d-validity}

Eq.~\eqref{eq:1d-effective} is exact on the curve axis, but a realistic nanowire possesses a finite transverse cross-section, over which the gauge-transformed tangential component varies.
While the generalized Poincar\'{e} gauge removes the transverse vector potentials identically on the axis, $\tilde{A}'_2 = \tilde{A}'_3 = 0$, a rigorous Taylor expansion of the transformed tangential component yields first-order linear deviations,
\begin{equation}
    \label{eq:taylor-1d}
    \tilde{A}'_1 \approx \tilde{A}_1(q^1, 0, 0) - q^2 \tilde{B}_3(q^1, 0, 0) + q^3 \tilde{B}_2(q^1, 0, 0),
\end{equation}
so that the transverse components of the effective magnetic field enter the tangential component.
Their projection onto the transverse ground state rests on spatial parity: the even ground state has $\langle q^2 \rangle = \langle q^3 \rangle = 0$, so that the first-order transverse magnetic coupling cancels identically,
\begin{equation}
    \langle \tilde{A}'_1 \rangle_\perp = \tilde{A}_1(q^1, 0, 0).
\end{equation}
Two conditions, both established in Appendix~\ref{app:bounds}, delimit the validity of this cancellation.
First, a transverse force $F_\perp$ gives the wavefunction a finite mean transverse position, $\langle q^2 \rangle, \langle q^3 \rangle \approx F_\perp d_\mathrm{rms}^2/\Delta E_\perp$, thereby destroying the definite parity of the ground state.
The mechanism survives as long as the induced energy shift remains negligible against the excitation gap,
\begin{equation}
    F_\perp d_\mathrm{rms} \ll \Delta E_\perp .
\end{equation}
The effective scalar potential is evaluated on the axis for the same reason: its linear transverse variation acts as the transverse force $-Q\nabla_\perp\tilde{\varphi}$, which displaces the transverse ground state and shifts $\tilde{\varphi}$ by $Q\langle q^\alpha\rangle\partial_\alpha\tilde{\varphi} = -Q^2|\nabla_\perp\tilde{\varphi}|^2d_\mathrm{rms}^2/\Delta E_\perp$, of second order in the transverse variation, the term $\frac{1}{2}Qd_\mathrm{rms}^2\nabla_\perp^2\tilde{\varphi}$ being of the same order.
Second, the adiabatic separation requires the transverse structure to remain intact, $l_B = \sqrt{\hbar c/|QB|} \gg d_\mathrm{rms}$.
The diamagnetic contribution scales as $B^2 q_\perp^2$, which is even under parity and hence does not violate the selection rule, but it restructures the transverse levels into Landau-like orbitals once $l_B$ becomes comparable to $d_\mathrm{rms}$, superseding the thin-wire suppression noted in Sec.~\ref{subsec:1d-formalism}.
The finite cross-section is therefore not an imperfection of the idealization but the very element that determines which symmetries survive in the tangential problem.

\subsection{Application to a corrugated quantum ring in a static magnetic field}\label{subsec:1d-ring}

To demonstrate the 1D formulation, we investigate a corrugated quantum ring lying in the plane perpendicular to the rotation axis $\bm{\omega}$, as shown in Fig.\ \ref{fig:applications} $(b)$.
The ring is subjected to the same laboratory-static field $\bm{B} = B_0\hat{\bm{x}} + B_z\hat{\bm{z}}$, with $B_0$ now lying in the plane of the ring.
Unlike a perfect circular ring, the corrugated geometry features a periodic modulation of the same local radial distance from the rotation axis, $\rho(\xi) = R + A\sin(N\xi)$, where $R$ is the base radius, $A$ the modulation amplitude, and $N$ the number of spatial periods.
In the small-corrugation limit ($\epsilon \equiv A/R \ll 1$ and $\epsilon N^2 \ll 1$), a Taylor expansion gives the local curvature fluctuation $k(\xi) \approx (1/R)\left[1 + \epsilon(N^2-1)\sin(N\xi)\right]$.
According to Eq.~\eqref{eq:dacosta-1d}, the corrugation generates a periodic 1D geometric potential
\begin{equation}
    V_\mathrm{g}^\text{1D}(\xi) \approx -\frac{\hbar^2}{8mR^2}\left[ 1 + 2\epsilon(N^2-1)\sin(N\xi) \right].
\end{equation}
Since $V_\mathrm{g}^\text{1D}$ is negative definite and quadratic in $k(\xi)$, it drives the particle toward the regions of largest curvature, i.e. toward the crests of the corrugation.
Depending on $k(\xi)$ alone, it is independent of both the field and the rotation. Throughout this subsection the scalar potential is held at zero, as in configuration (B) of Sec.~\ref{sec:three-dimensional}.
Because the curve is confined to the $z = 0$ plane, its tangent vector has no $z$ component, and the in-plane field generates the vector potential $\bm{A}_{B_0} = (B_0/2)\,\rho(\xi)\sin(\xi + \omega t)\,\hat{\bm{z}}$, whose tangential projection vanishes identically, $\tilde{A}_{B_0,\xi} = 0$.
The in-plane field nevertheless enters the gauge-transformed tangential potential $\tilde{A}'_1$ linearly in $q^2$ and $q^3$ [Eq.~\eqref{eq:taylor-1d}] and is thereby removed by the parity selection rule of Sec.~\ref{subsec:1d-validity}, so that the effective 1D equation is insensitive to $B_0$ to first order in the field.
In contrast to the torus, where the poloidal component $\tilde{A}_\theta$ no longer vanishes for $B_0 \neq 0$ and the axial symmetry is thereby broken (Sec.~\ref{subsec:2d-torus}), no zero-order tangential component exists on a planar curve: the in-plane field enters only through the odd-parity term of Eq.~\eqref{eq:taylor-1d}, whose linear response is annihilated by spatial parity.
The tangential dynamics is thus governed solely by the axial field and the rotation, which combine into the effective field $B_\mathrm{eff} = B_z + 2mc\omega/Q$.
For the effective fields, the covariant azimuthal component of the Sagnac vector potential and the centrifugal scalar potential are approximated as
\begin{equation}
\begin{split}
    \tilde{A}_\xi(\xi) &\approx \frac{1}{2}B_\mathrm{eff} R^2\left[ 1 + 2\epsilon\sin(N\xi) \right], \\
    V_\mathrm{cf}(\xi) &\approx -\frac{1}{2}m\omega^2 R^2 \left[ 1 + 2\epsilon\sin(N\xi) \right].
\end{split}
\end{equation}
The angular metric is $G_{\xi\xi} = \rho^2 + (\partial_\xi\rho)^2$, approximated to first order in $\epsilon$ by $R^2\left[1 + 2\epsilon\sin(N\xi)\right]$.
Substituting these into Eq.~\eqref{eq:1d-effective} yields the 1D Schr\"{o}dinger equation governed by the full Laplace--Beltrami operator:
\begin{equation}
    \mathrm{i}\hbar\,\partial_t \chi_\mathrm{t} = \biggl[ \frac{1}{2m}\bigl( G_{\xi\xi}^{-1/2} \hat{\Pi}_\xi \bigr)^{2} + V_\mathrm{g}^\text{1D}(\xi) + V_\mathrm{cf}(\xi) \biggr]\,\chi_\mathrm{t},
\end{equation}
where $\hat{\Pi}_\xi = -\mathrm{i}\hbar\,\partial_\xi - (Q/c)\tilde{A}_\xi(\xi)$.
For $B_z = 0$, the diamagnetic contribution cancels $V_\mathrm{cf}(\xi)$ up to corrections of $\mathcal{O}(\epsilon^2)$; for $B_z \neq 0$, this cancellation is superseded by the same joint Sagnac--Aharonov--Bohm residual as in Eq.~\eqref{eq:residual-Bz}, with $\theta$ replaced by $\xi$.
Extracting the constant background, the geometric variation together with the periodic flux modulation $\delta\tilde{A}_\xi(\xi) = B_\mathrm{eff}RA\sin(N\xi)$ produces an artificial superlattice.
We find that, under periodic spatial modulations, the 1D geometric curvature acts as an intrinsic scalar grating that generates Bloch bands, leaving the Sagnac phase structurally stable.
The corrugated ring therefore combines an axial-field-tunable superlattice with an in-plane-field immunity that is protected by the parity selection rule rather than by the smallness of the field.

The 1D construction is complete at this point. The Bishop frame removes the torsion-induced frame rotation, so that Eq.~\eqref{eq:1d-effective} carries no trace of the Frenet--Serret singularity; the first-order transverse magnetic coupling that survives the gauge is annihilated by the same parity selection rule as in the 2D case, under the two conditions $F_\perp d_\mathrm{rms} \ll \Delta E_\perp$ and $l_B \gg d_\mathrm{rms}$. The corrugated ring shows what this yields: the periodic curvature acts as an intrinsic scalar grating which, together with the periodic flux modulation, generates Bloch bands while leaving the Sagnac phase structurally stable, and the planar geometry leaves no zeroth-order tangential component for the in-plane field to act on, so that its first-order effect vanishes. Set against the torus, where the in-plane field breaks the axial symmetry, the same field is here filtered out: the response is governed by the out-of-plane extent of the structure rather than by its dimensionality.

\section{Conclusions}\label{sec:conclusion}

The above analysis reduces the quantum dynamics of a charged particle on a rotating curved manifold of arbitrary shape to two ingredients---an effective gauge potential and a geometric potential---once the non-relativistic reduction of the covariant Klein--Gordon equation is combined with thin-layer quantization.
The residual gauge--geometry couplings are eliminated by three mechanisms, two of which remove artifacts of the description and one of which removes a genuine coupling: a generalized transverse Poincar\'{e} gauge, which removes the transverse component of the vector potential identically on the reference manifold; the rotation-minimizing Bishop frame, which removes the torsion-induced frame rotation; and the spatial parity of the transverse ground state, which annihilates the first-order transverse magnetic coupling that survives both the gauge choice and the frame change, in either geometry. The parity argument requires a single, inversion-symmetric transverse well, and the adiabatic separation requires the transverse level spacing to exceed every other scale of the problem. Beyond these bounds the cancellation is no longer exact, and the residual coupling is suppressed by the small dimensionless ratio $F_\perp d_\mathrm{rms}/\Delta E_\perp$---the mean transverse displacement measured in units of $d_\mathrm{rms}$---while the transverse width must also be small compared with the magnetic length, $l_B \gg d_\mathrm{rms}$.
The two model geometries show that this reduction is not merely formal. In the absence of a transverse field the two contributions separate exactly, the effective vector potential fixing the interference phase and the da Costa potential the amplitude profile.
Periodic geometric and flux modulations turn the corrugated ring into an artificial superlattice whose Bloch bands are tuned by the axial field, while on the torus the axial field tunes the curvature-induced localization through the joint Sagnac--Aharonov--Bohm residual of Eq.~\eqref{eq:residual-Bz}. What distinguishes the two cases is not the dimension but the out-of-plane extent: the in-plane field breaks the axial symmetry of the curved torus, whereas on the planar ring it has no tangential component at zeroth order and its residual first-order coupling is annihilated by the parity selection rule.

The framework as it stands is restricted to a single transverse well, to spinless particles, to the thin-layer regime, and to external sources that hold the rotating-frame scalar potential at zero; it is precisely these restrictions that make the decoupling exact rather than approximate.
A natural extension of this formalism is the incorporation of intrinsic spin \cite{matsuo2011spin}.
By starting from the covariant Dirac equation rather than the KG equation, future studies can investigate the interplay between the curvature-induced Rashba spin-orbit interaction and the non-inertial spin-rotation coupling $-\bm{\omega} \cdot \hat{\bm{S}}$, where $\hat{\bm{S}}$ denotes the spin operator \cite{strange2016dirac}.
Such an extension would show how the non-inertial spin-rotation coupling competes with the curvature-induced Rashba term in fixing the spin texture of a rotating nanotube, and would thereby provide a basis for modelling spin-mechatronic devices, including dynamically rotating carbon nanotubes and nano-electromechanical systems operating in the extreme quantum limit.

\begin{acknowledgments}
This work is jointly supported by the National Natural Science Foundation of China (Grant No.~12475019), the Natural Science Foundation of Shandong Province (Grant No.~ZK2024MA038), and the National Laboratory of Solid State Microstructures at Nanjing University (Grant No.~M39059).
\end{acknowledgments}

\appendix

\section{Order-of-magnitude estimates for the transverse validity bounds}\label{app:bounds}

All bounds quoted in Secs.~\ref{subsec:2d-validity} and \ref{subsec:1d-validity} follow from perturbation theory for a transversely confined particle, carried to the leading non-vanishing order in the transverse force.
Throughout this appendix the transverse degrees of freedom are denoted collectively by $\bm{q}_\perp$ ($q^3$ for surfaces and $\{q^2, q^3\}$ for curves), and the particle is taken to be localized by a confining well $V_\mathrm{c}(\bm{q}_\perp)$ of otherwise arbitrary form; no particular functional dependence is assumed, and none of the estimates below depends on one.
Apart from the transverse isotropy assumed below, the only property required of $V_\mathrm{c}$ is that its transverse ground state be non-degenerate, for the reason given below.
Only two ground-state properties of the well enter, namely the transverse level spacing $\Delta E_\perp = E_1 - E_0$ between the first excited and the ground transverse states, and the per-direction mean-square width $d_\mathrm{rms}^2 \equiv \langle (q^\alpha)^2 \rangle_0$; the well is taken to be isotropic in the transverse plane, so that $\langle q^\alpha q^\beta \rangle_0 = d_\mathrm{rms}^2\,\delta^{\alpha\beta}$.
Since $d_\mathrm{rms}$ fixes the kinetic-energy scale of the ground state, these two quantities are linked, for a single well whose transverse ground state is non-degenerate, by the estimate $\Delta E_\perp \sim \hbar^2/(2md_\mathrm{rms}^2)$, which is used below only to trade one for the other whenever a numerical reduction is required.
A well with a (nearly) degenerate transverse ground state---a symmetric double well, for instance, whose $\Delta E_\perp$ is set by tunnelling and may be exponentially small while $d_\mathrm{rms}$ remains large---falls outside the adiabatic assumptions of the thin-layer reduction and is not considered here.
Because the resulting relations are order-of-magnitude criteria, numerical factors of order unity are not retained.

A uniform transverse force $\bm{F}_\perp$ enters as the perturbation $\hat{H}' = -\bm{F}_\perp \cdot \bm{q}_\perp$.
Such a force may equally originate from the transverse gradient of the effective scalar potential, $\bm{F}_\perp = -Q\nabla_\perp\tilde{\varphi}$, so that the evaluation of $\tilde{\varphi}$ on the reference manifold in Eqs.~\eqref{eq:2d-effective} and~\eqref{eq:1d-effective} is subject to the same bound.
The reference manifold is chosen to pass through the centre of the ground-state transverse density, $\langle q^\alpha \rangle_0 = 0$, so that the first-order shift $-\bm{F}_\perp\cdot\langle \bm{q}_\perp \rangle_0$ vanishes identically and the response is governed by second-order perturbation theory,
\begin{equation}
\begin{split}
    \Delta E_0 &= -\sum_{n\neq0} \frac{\bigl|\langle n|\bm{F}_\perp\cdot\bm{q}_\perp|0\rangle\bigr|^2}{E_n - E_0},\\
    |\delta \chi_0\rangle &= \sum_{n\neq0} \frac{\langle n|\bm{F}_\perp\cdot\bm{q}_\perp|0\rangle}{E_0 - E_n}\,|n\rangle ,
\end{split}
\end{equation}
where $\Delta E_0$ is the ground-state energy shift and $|\delta \chi_0\rangle$ the first-order correction to the ground-state wavefunction.
By the closure relation the squared matrix elements sum to $\sum_n |\langle n|\bm{F}_\perp\cdot\bm{q}_\perp|0\rangle|^2 = |\bm{F}_\perp|^2 d_\mathrm{rms}^2$, whereas every energy denominator is bounded from below by $\Delta E_\perp$.
Evaluating the denominators at $\Delta E_\perp$ thus reproduces the energy shift and the polarization of the transverse density up to the factor of order unity dropped throughout this appendix,
\begin{equation}
    \Delta E \simeq -\frac{|\bm{F}_\perp|^2 d_\mathrm{rms}^2}{\Delta E_\perp},
    \quad
    \bigl|\langle \bm{q}_\perp \rangle\bigr| \simeq \frac{|\bm{F}_\perp| d_\mathrm{rms}^2}{\Delta E_\perp},
\end{equation}
where $\langle \bm{q}_\perp \rangle$ is the mean position acquired by the transverse wavefunction through the admixture $|\delta \chi_0\rangle$.
Requiring this lowering to remain small compared with the level spacing, $|\Delta E| \ll \Delta E_\perp$, is equivalent to
\begin{equation}
    \frac{|\bm{F}_\perp| d_\mathrm{rms}}{\Delta E_\perp} \simeq \frac{\bigl|\langle \bm{q}_\perp \rangle\bigr|}{d_\mathrm{rms}} \ll 1 ,
\end{equation}
which states that the work done by the force across the width $d_\mathrm{rms}$ of the transverse ground state stays well below the transverse excitation energy, or equivalently that the induced displacement is small compared with that width.
Both forms involve only $\Delta E_\perp$ and $d_\mathrm{rms}$, and hence hold for any single-well potential $V_\mathrm{c}$.

The same perturbative scheme quantifies the accuracy of evaluating the effective potentials on the reference manifold.
The residual tangential component introduced in Sec.~\ref{subsec:2d-validity}, $\delta\tilde{A}'_a = q^3\,(\partial_3 \tilde{A}_a - \partial_a \tilde{A}_3)$, couples to first order only through the odd-parity coordinate $q^3$ and therefore averages to zero for a symmetric ground state.
Its leading surviving contribution is second order and even, giving a relative correction to the tangential kinetic energy of order
\begin{equation}
    \left[ \frac{d_\mathrm{rms}\,\bigl(\partial_3 \tilde{A}_a - \partial_a \tilde{A}_3\bigr)}{\tilde{A}_a} \right]^2 \ll 1 ,
\end{equation}
so that the on-surface form of the effective equation is accurate as long as $d_\mathrm{rms}|\partial_3 \tilde{A}_a - \partial_a \tilde{A}_3| \ll |\tilde{A}_a|$.

Finally, the diamagnetic term $Q^2B^2|\bm{q}_\perp|^2/(8mc^2)$ shifts the transverse levels by ${\sim}\,Q^2B^2\langle|\bm{q}_\perp|^2\rangle_0/(8mc^2) = \mathcal{N}\hbar^2 d_\mathrm{rms}^2/(8m l_B^4)$, where $\mathcal{N} = 1$ for a surface and $\mathcal{N} = 2$ for a curve counts the transverse directions, so that $\langle|\bm{q}_\perp|^2\rangle_0 = \mathcal{N} d_\mathrm{rms}^2$, and $l_B = \sqrt{\hbar c/|QB|}$ is the magnetic length.
Requiring this shift to remain well below the actual excitation gap, $\mathcal{N}\hbar^2 d_\mathrm{rms}^2/(8m l_B^4) \ll \Delta E_\perp$, yields the model-independent bound
\begin{equation}
    l_B^4 \gg \frac{\mathcal{N}\hbar^2 d_\mathrm{rms}^2}{8m\,\Delta E_\perp},
\end{equation}
which retains the true transverse gap rather than the width alone.
Only for a typical single well, where $\Delta E_\perp \sim \hbar^2/(2md_\mathrm{rms}^2)$, does this bound reduce---once more up to factors of order unity---to the purely geometric condition $l_B \gg d_\mathrm{rms}$.

The same estimates carry over verbatim to the 1D case of Sec.~\ref{subsec:1d-validity}, with $\bm{q}_\perp = (q^2, q^3)$ denoting the transverse coordinates.
Since $\langle (q^2)^2 \rangle_0 = \langle (q^3)^2 \rangle_0 = d_\mathrm{rms}^2$, the relative correction derived above applies equally to the residual component of Eq.~\eqref{eq:taylor-1d}.
If the parity of the transverse ground state is lifted by a transverse force, the strict cancellation $\langle \tilde{A}'_1 \rangle_\perp = \tilde{A}_1(q^1, 0, 0)$ is replaced by a residual of order
\begin{equation}
    \big\langle \tilde{A}'_1 \big\rangle_\perp - \tilde{A}_1(q^1, 0, 0) \sim \frac{|\bm{F}_\perp| d_\mathrm{rms}^2}{\Delta E_\perp}\,\bigl|\tilde{\bm{B}}_\perp\bigr| ,
\end{equation}
which vanishes in the strict thin-wire limit and remains negligible as long as $|\bm{F}_\perp| d_\mathrm{rms} \ll \Delta E_\perp$.
Finally, in this case $l_B \gg d_\mathrm{rms}$ ensures that the transverse bound states retain their width $d_\mathrm{rms}$ rather than being restructured into Landau-like orbitals.

\section{The transverse Poincar\'{e} gauge condition in the normal plane}\label{app:poincare-radial}

This appendix supplies the algebra behind the gauge condition quoted in Sec.~\ref{subsec:1d-formalism}, and shows that the linear form \eqref{eq:residual-transverse} follows from that condition together with the field strength on the axis.
Greek indices run over the transverse pair, $\alpha,\beta = 2,3$, the summation convention is understood, and the axis $q^2 = q^3 = 0$ is the reference manifold.
The gauge function of Eq.~\eqref{eq:gauge-1d} depends on the transverse coordinates both through the explicit factor $q^\alpha$ and through the rescaled argument $\lambda\bm{q}_\perp$; since $q^1$ is not rescaled, differentiation with respect to a transverse coordinate gives
\begin{equation}
\label{eq:dgamma}
    \partial_\beta\gamma = -\int_0^1 \Bigl[ \tilde{A}_\beta(\lambda \bm{q}_\perp) + \lambda\, q^\alpha\, \partial_\beta \tilde{A}_\alpha(\lambda\bm{q}_\perp) \Bigr] \mathrm{d}\lambda .
\end{equation}
The two terms originate from these two dependences: $\partial_\beta q^\alpha = \delta^\alpha_\beta$ replaces the summed index by $\beta$ in the first, whereas the chain rule $\partial_\beta\tilde{A}_\alpha(q^1,\lambda q^2,\lambda q^3) = \lambda\,(\partial_\beta\tilde{A}_\alpha)(\lambda\bm{q}_\perp)$ produces the second, the parameter $\lambda$ itself being independent of $q^\beta$.
Contracting Eq.~\eqref{eq:dgamma} with $q^\beta$ turns the integrand into a total derivative in $\lambda$,
\begin{equation}
    q^\beta \tilde{A}_\beta(\lambda\bm{q}_\perp) + \lambda\, q^\alpha q^\beta\, \partial_\beta \tilde{A}_\alpha(\lambda\bm{q}_\perp)
    = \frac{\mathrm{d}}{\mathrm{d}\lambda}\bigl[\lambda\, q^\alpha \tilde{A}_\alpha(\lambda\bm{q}_\perp)\bigr],
\end{equation}
so that the integral is fixed by its boundary values alone,
\begin{equation}
\label{eq:dgamma-contracted}
    q^\beta\partial_\beta\gamma = -\int_0^1 \frac{\mathrm{d}}{\mathrm{d}\lambda}\bigl[\lambda\, q^\alpha \tilde{A}_\alpha(\lambda\bm{q}_\perp)\bigr]\mathrm{d}\lambda = -q^\alpha\tilde{A}_\alpha(\bm{q}_\perp).
\end{equation}
The gauge-transformed potential $\tilde{A}'_\alpha = \tilde{A}_\alpha + \partial_\alpha\gamma$ therefore obeys
\begin{equation}
\label{eq:radial-condition}
    q^\alpha\tilde{A}'_\alpha = q^\alpha\tilde{A}_\alpha + q^\alpha\partial_\alpha\gamma = 0
\end{equation}
identically, which is the condition quoted in Sec.~\ref{subsec:1d-formalism}.

Eq.~\eqref{eq:radial-condition} is the transcription to a two-dimensional transverse plane of the surface gauge of Eq.~\eqref{eq:poincare-2d}: when the transverse space is one-dimensional the contraction involves a single component and the identity collapses to $\tilde{A}'_3 = 0$.
It is a single scalar condition, and no gauge function can impose more than one on the two transverse components: it states that $\tilde{A}'_\perp$ carries no radial part along the rays issuing from the axis, leaving the azimuthal part, which is the gauge-invariant content of the transverse field, untouched.
Equivalently $\tilde{A}'_\alpha = \varepsilon_{\alpha\beta}q^\beta\Phi(\bm{q}_\perp)$ for some function $\Phi$.
Because $\partial_\alpha\gamma|_{\bm{q}_\perp=0} = -\tilde{A}_\alpha(0)$, the transformed components vanish on the axis; and since they must remain regular there, the only gauge functions compatible with Eq.~\eqref{eq:radial-condition}, namely those obeying $q^\alpha\partial_\alpha\Lambda = 0$, are constant on each transverse plane and leave $\tilde{A}'_\alpha$ unchanged.
The transverse gauge is therefore fixed completely.

Expanding about the axis as $\tilde{A}'_\alpha = \mathcal{M}_{\alpha\beta}q^\beta + O(q_\perp^2)$, Eq.~\eqref{eq:radial-condition} requires $q^\alpha \mathcal{M}_{\alpha\beta} q^\beta = 0$ for arbitrary $\bm{q}_\perp$, that is, $\mathcal{M}_{\alpha\beta} = -\mathcal{M}_{\beta\alpha}$: the gauge condition supplies the antisymmetry.
In two dimensions such a matrix has a single independent entry, $\mathcal{M}_{23}$ (with $\mathcal{M}_{32} = -\mathcal{M}_{23}$), which the tangential field on the axis fixes through
\begin{equation}
    \tilde{B}_1 \equiv \bigl(\partial_2\tilde{A}'_3 - \partial_3\tilde{A}'_2\bigr)_{q^2=q^3=0} = -2\mathcal{M}_{23} ,
\end{equation}
whence $\mathcal{M}_{23} = -\tilde{B}_1/2$.
The gauge condition thus fixes the symmetry of the transverse potential while the field strength fixes its scale, which is the origin of the symmetric-gauge form \eqref{eq:residual-transverse}.

Finally, substitution into the transverse part of the minimal coupling reproduces the terms of Eq.~\eqref{eq:1d-transverse-hamiltonian}: since $\nabla_\perp\cdot\tilde{\bm{A}}'_\perp = 0$ for the constant field on the axis, the linear terms give the Zeeman-like coupling
\begin{equation}
    \frac{\mathrm{i}\hbar Q}{mc}\,\tilde{A}'_\alpha\partial_\alpha = -\frac{Q\tilde{B}_1}{2mc}\,\hat{L}_1 ,
    \qquad \hat{L}_1 = -\mathrm{i}\hbar\,(q^2\partial_3 - q^3\partial_2),
\end{equation}
while the quadratic terms give the diamagnetic contribution $Q^2\tilde{B}_1^2|\bm{q}_\perp|^2/(8mc^2)$.

\bibliography{refs}

\end{document}